# *Articulation entre élaboration de solutions et argumentation polyphonique*

**Michael Baker***

**Françoise Détienne****

**Kristine Lund***

**Arnauld Séjourné***

\* GRIC — CNRS / Université Lyon 2
\*\* Projet Eiffel " Cognition et coopération en conception ", INRIA-Rocquencourt

RESUME
Cette analyse tente d'une part de montrer l'articulation entre l'expression des arguments exprimés en (dé)faveur de la demande de modifications et de la solution existante et la co-élaboration de nouvelles solutions, et d'autre part le rôle du processus de la polyphonie (« plusieurs voix »). …
MOTS-CLES
Pragma-dialectique, ergonomie de la conception, argumentation polyphonique

## 1. INTRODUCTION

Cette recherche vise l'étude de la dynamique de l'action collective dans une situation de re-conception en architecture, grâce à une collaboration pluridisciplinaire entre des chercheurs en ergonomie cognitive, sciences du langage et didactique des sciences. L'objectif de recherche principal est d'élaborer une méthode d'analyse de l'interaction communicative entre des concepteurs, appréhendée dans ses dimensions langagières et gestuelles, qui serait généralisable au-delà de la situation d'étude et pertinente pour la conception des dispositifs d'aide à la conception.

Dans la situation d'étude, trois architectes de la société « W » (à Paris), ayant des rôles et compétences différents, sont réunis afin de collaborer dans la re-conception d'un plan architectural d'aménagement d'un château en centre de séminaires, en réponse à des demandes de modifications écrites, formulées par un représentant de la propriétaire du bâtiment.

Par définition, toute situation de conception, compte tenu de sa dimension dynamique, implique une évolution conjointe de la représentation du problème et des solutions élaborées (Falzon et al. 1990), Dans le cas présent nous insistons sur cet aspect dans la mesure où il s'agit de modifier une conception déjà réifiée dans un plan, en réponse à des demandes provenant de l'extérieur de l'équipe de conception, au sens étroit du terme. L'activité de re-conception survenant à la demande des commanditaires du projet a lieu dans une classe plus large de situations de conception collective.

Dans cet article nous nous centrons sur un problème particulier et complexe, celui de la diminution éventuelle du nombre d'ascenseurs dans le bâtiment. Ce problème mobilise à la fois des notions de circulation sur les plans verticaux et horizontaux, et des notions de fonctionnalités et d'espace en interaction avec des contraintes de coût et d'esthétique. Notre analyse tentera d'une part à montrer l'articulation entre l'expression des arguments exprimés en (dé)faveur de la demande de modifications et de la solution existante et la co-élaboration de nouvelles solutions, et d'autre part le rôle du processus de la polyphonie (« plusieurs voix »). Dans ce cas, la polyphonie correspond à la reconstruction, dans l'interaction argumentative entre architectes, du point de vue des « absents » (la propriétaire, son gestionnaire).

## 2 LA SITUATION D'ETUDE

Notre analyse prend comme point de départ une analyse *a priori* (ou, une analyse dite « rationnelle ») et statique de la situation-tâche à réaliser en commun par les trois architectes, sur laquelle va se greffer une analyse des processus séquentiels et dynamiques (voir la Figure 1).

L'interaction s'articule autour de la mise en relation de deux supports tangibles : le plan architectural, *P*, élaboré dans la société W, et le fax, *F*, comportant une liste de demandes, *D*, de modifications. Au centre de l'activité collective se trouve une **délibération** : *dans quelle mesure* *prendra-t-on en compte la demande $D_n$, vis-à-vis P* ? En premier lieu, cette prise de décision requiert une activité de **problématisation** (pour décider entre x et y il faut se mettre d'accord sur leurs significations), qui se décompose en une activité de **reconstruction** double : quel est le point de vue (motivation, arguments, critères, …) sous-jacent à la demande $D_n$, et quel était « notre » (i.e. celui des architectes présents dans la situation d'étude) point de vue sous-jacent à l'élément ɸ évoqué par $D_n$ ? Ces reconstructions se traduiront par une négociation du sens dans l'interaction. Le point de vue des absents, représentés par le support écrit, sera, bien entendu, étudié du point de vue des architectes présents. Dans le cas étudié ici, $D_n$ peut se paraphraser par « diminuer le nombre d'ascenseurs » et correspond aux ascenseurs qui se situent à côté du restaurant.

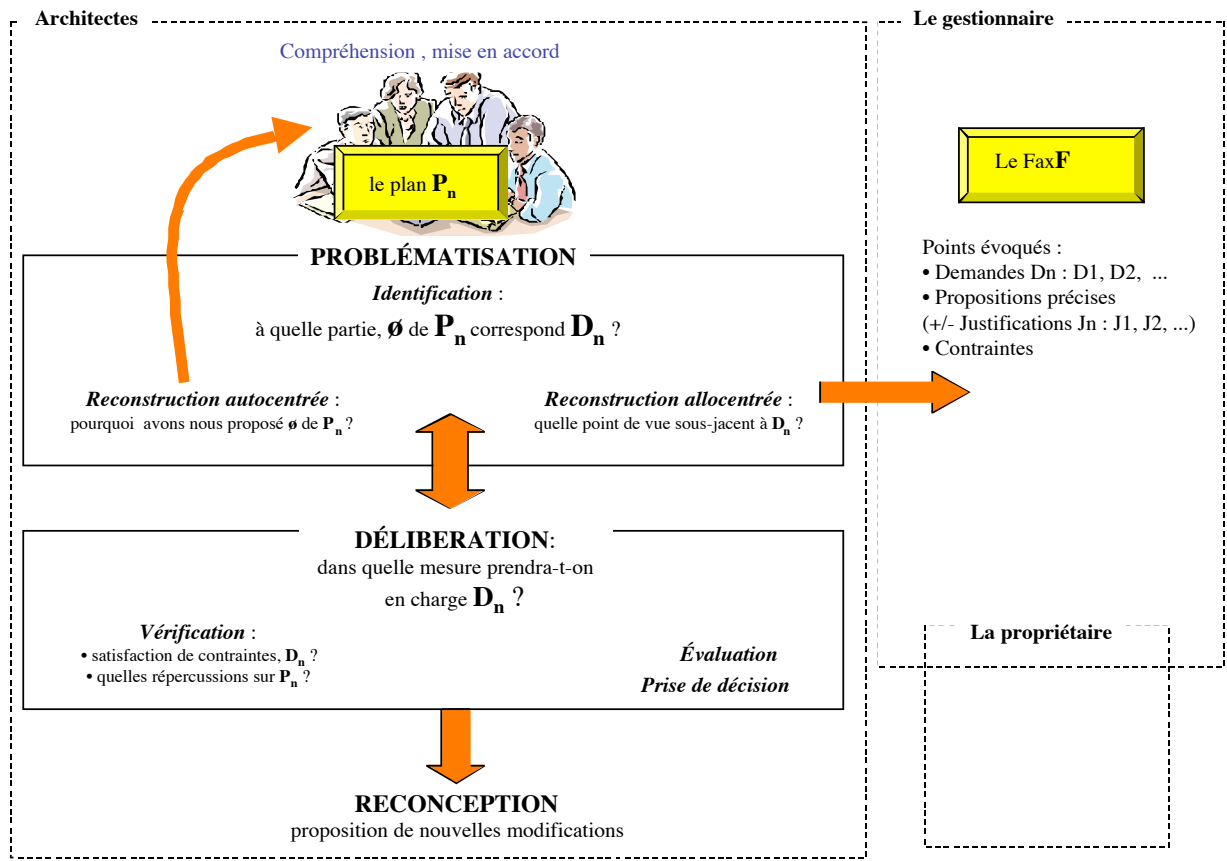

*Figure 1. Analyse à priori de la situation de reconception*

La délibération même, fondée sur des reconstructions jugées adéquates dans le contexte de l'interaction, s'effectuera par l'évocation d'arguments, qui permettront de vérifier si des contraintes multiples seraient satisfaites et ainsi de fonder des évaluations des solutions proposées.

Dans le cas où la demande serait jugée recevable, la confrontation entre élément du plan et demande de modification sera éliminée par la re-conception, c'est-à-dire, par la co-élaboration de nouvelles solutions architecturales, pour satisfaire de multiples contraintes « des deux côtés ».

Si ces sous-tâches sont séparables d'un point de vue analytique, elles entretiendront des interactions multiples dans la dynamique de l'activité : par exemple, il y aurait un va-et-vient entre reconstruction, évaluation et co-élaboration. C'est cette « interaction » entre éléments de l'activité qu'il s'agit d'analyser dans l'interaction communicative.

## 3  METHODE D'ANALYSE

Les éléments de l'activité cognitive de re-conception sont associés à des activités discursives spécifiques, mises en œuvre dans l'interaction communicative.

*Délibération, argumentation et processus de re-conception*

La délibération portant sur l'élément du plan et la demande de modification sera analysée en tant qu'une interaction argumentative délibérative. Selon la typologie de Walton (1989), il s'agit d'une argumentation de type « légal » (forensic), orientée par l'examen « objectif » et coopératif des différentes hypothèses, à la lumière des indices et des arguments, au lieu d'un débat entre adversaires avoués. Cependant, dans la présente situation d'étude, un tel type d'interaction argumentative doit être analysé en tenant compte de ses interactions avec la résolution du problème : celle-ci, quand elle est effectuée collectivement, peut engendrer de nouvelles questions à débattre, et les thèses issues des débats fournissent de nouvelles données pour la résolution.

Pour ces raisons, nous fondons notre démarche sur une méthode d'analyse déjà élaborée pour l'étude des interactions argumentatives produites en situations de résolution de problèmes, en l'occurrence, il s'agit des situations d'apprentissage en sciences (Baker, 1999 ; 2002 ). Cette démarche est « pragma-dialectique » (Barth & Krabbe, 1982 ; van Eemeren & Grootendorst, 1984)1, dans la mesure où des types d'interventions argumentatives (attaques, défenses, concessions, retractions, … portant sur des thèses) sont produites en vue de déterminer l'issu d'un jeu de dialogue. La démarche est étendue par la prise en compte des dimensions suivantes2 :

— *la dimension interactive* ; il s'agit des processus de négociation du sens des enjeux du débat, des reformulations des propositions de solutions, de la co-élaboration de connaissances et de solutions nouvelles ; une typologie des transformations sur le plan des connaissances (par ex. « généralisation, particularisation, élaboration, restriction, …) peut être proposée (Baker, 1994) ;
— *la dimension épistémologique* ; il s'agit de la nature des connaissances évoquées en tant qu'arguments, qui ont, par voie de conséquence, des « poids » et des ancrages cognitifs différents ;
— *la dimension notionnelle* ; il s'agit de la manière de reconceptualiser le domaine de référence, introduisant, par exemple, de nouvelles associations et dissociations notionnelles (c.f. Perelman & Olbrechts-Tyteca, 1988 [1958]).

En s'appuyant sur ces dimensions supplémentaires, il est possible de démontrer comment l'argumentation dialoguée peut se « dissoudre » (au lieu de se « résoudre ») par la co-élaboration d'une nouvelle solution (dimension 1), à partir des différentes propositions-thèses, et la reconceptualisation du problème même, et le rôle des différents types d'arguments, s'appuyant sur des contraintes en conception architecturale, dans la prise de décision argumentative.

Toutes ces dimensions additionnelles prennent substance en référence avec un modèle ergonomique de la conception (Darses, Détienne & Visser, 2001), appliqué au raisonnement architectural, qui distingue notamment :

— des attributs des couples problèmes/solutions, comme la « circulation verticale », la « circulation horizontale », les « fonctionnalités » des espaces. Une forte cohérence entre le choix des valeurs pour ces attributs est recherchée (par ex. le « Besoin d'un salon pour l'attente/arrivée-continuité des espaces ») notamment à travers la satisfaction de contraintes (par ex. esthétique) ;
— des contraintes prescrites ou inférées/ajoutées au cours du raisonnement architectural, comme le coût et l'esthétique ; ils apparaîtront sous la dimension épistémologique de l'argumentation.

*L'argumentation, reconstruction des points de vue et polyphonie*

Notre deuxième axe d'analyse, étroitement lié au premier, tente d'intégrer la dimension polyphonique (« plusieurs voix ») de l'interaction argumentative, grâce à la notion du « discours rapporté » (ici, celui du propriétaire du bâtiment, qui lui, passe par la voix de son représentant, le gestionnaire) : « Le discours rapporté, c'est le *discours dans le discours*, l'*énonciation dans l'énonciation*, mais c'est, en même temps, un *discours sur le discours*, une *énonciation sur l'énonciation.* » (Bakhtine, 1977 [1929], p. 161 ; italiques de l'auteur).

Ce discours rapporté, *a priori* « figé » par le support écrit (le fax), et à reconstruire par les architectes participants, joue le rôle d'un participant absent dans l'interaction argumentative. Nous postulons que les architectes, partants d'une lecture à haute voix du discours du gestionnaire, s'éloigneront progressivement de ceci par raffinements successifs ; ils diminueront en ce faisant la « distance » entre le discours effectif et le discours rapporté sur le plan de la prise en charge dans l'énonciation. Il est donc nécessaire, dans notre analyse, non seulement d'analyser le discours échangé en situation, mais également de représenter l'évolution progressive de ce discours dans le discours.

---

[1] Pour une introduction succincte en langue française on peut se rapporter à Heinzmann (1992).

[2] Le grand absent, bien entendu, de ces quelques dimensions est la dimension socio-cognitive, portant sur les enjeux relationnels de l'argumentation.

## 4  EXEMPLE d'UNE ANALYSE

L'analyse se présente sous la forme d'un tableau, comme suite (Tableau 1) :

TABLEAU 1. GRILLE D'ANALYSE

| L | Loc | Élaborations de solutions et Argumentation ||||||  Tâche / connaissances | Légende |
|---|---|---|---|---|---|---|---|---|---|
| | | Participants ||||||| |
| | | Marie (M) || Charles (C) || Louis (L) ||| |
| | | Élaboration | Argumentation | Élaboration | Argumentation | Élaboration | Argumentation | | |
| | | | | | | | | | |

Pour chaque participant dans la situation (M, C, et L), on distingue les élaborations des solutions (Pn) et les arguments (An) exprimés, avec leurs polarités (+, —) vis-à-vis des propositions, qui deviennent thèses une fois qu'il existe une diversité de types d'arguments (+ et —). De cette manière, on peut saisir l'articulation entre résolution de problèmes (élaboration progressive de solutions) et argumentation. La colonne « légende » précise de type d'élaboration ou d'extension des solutions effectuée, et la colonne « tâche / connaissances » précise les types de connaissances, contraintes, facteurs,… évoqués dans le raisonnement.

Le Tableau 2, ci-dessous, montre un court extrait du corpus d'étude, où les architectes commencent leur discussion/délibération sur la demande du gestionnaire, qui souhaite « diminuer le nombre d'ascenseurs », avec une première analyse des types d'arguments (dimension épistémologique).

TABLEAU 2. DEBUT DE LA DISCUSSION SUR LA DIMINUTION DU NOMBRE D'ASCENSEURS.

| Transcription | Types d'arguments |
|---|---|
| 642. C. alors / voilà \ parce que finalement on en arrive \ donc on a fait on a parlé du fitness agencement (…) il veut diminuer le nombre d'ascenseurs / | |
| 643. (..) | |
| 644. L. [((toux)) | |
| 645. C. [baisse des coûts d'aménagements et de: d'amé:: baisse du coût d'aménagements et des coûts d'entretien (…) proposition supprimer l'ascenseur restaurant (1.0) accès princi[pal | ARGT+ Coût |
| 646. M.            [(c'est ça l'ascenseur près de l'accueil) | |
| 647. C. par l'ascenseur côté salle de réunions \ supprimer l'ascenseur près de l'accueil [ah y a | |
| 648. L.            [mm mm | |
| 649. C. c'est là où il y a une confusion (..) parce que pour lui | |
| 650. (..) | |
| 651. M. on supprime ça / ou ça / | |
| 652. L. oui (4.0) enfin d'après lui moi [parce que | |
| 653. M.                [e::t | |
| 654. L. moi lui il pense que:: on peut fonctionner avec un seul euh ascenseur ou monte-charge ici (..) euh pour l'instant en fait nous ce qu'on avait prévu c'était (qu' t'avais) un accueil ici un petit peu privilégié avec un ascenseur client / | ARGT— double fonction pour ascenseur monte-charge<br>ARGT+ Utilisation (remarque : évaluation comparative avec existant: ARGT— pour P) |

L'analyse, sous la forme montrée dans le Tableau 1, distribuera les extensions des solutions, et les arguments qui s'y rattachent, au travers les participants. À chaque élément de l'analyse — proposition-thèse ou argument — on précise le point de vue évoqué de celui de l'énonciateur. Par exemple, un argument énoncé par C, mais qui rapporte le discours de G (le gestionnaire) sera noté dans la colonne pour C : « A$_{<G>}$ ». Cette analyse est reproduite dans le Tableau 3 ci-dessous. Pour des raisons de place, le contenu de la colonne « légende » est représentée sous le tableau. Les symboles de type {—} ou {+}, placés avant le contenu (P, A) représentent des opinions ou engagements des énonciateurs, distingués de la polarité de l'argument (par ex., dans la colonne pour l'énonciateur C, « {−}A<G> » représente le fait que C n'accepte pas l'argument A de G).

TABLEAU 3. ANALYSE ARGUMENTATIVE DE L'EXTRAIT.

| L | Loc | Élaborations de solutions et Argumentation ||||||Tâche / connaissances |
|---|---|---|---|---|---|---|---|---|
| | | Participants ||||||  |
| | | Marie (M) || Charles (C) || Louis (L) ||  |
| | | Élaboration | Argumentation | Élaboration | Argumentation | Élaboration | Argumentation |  |
| 642 | C | | | $P1_{<G>}$ | | | | C.V. |
| 643 | ? | | | | | | | |
| 644 | L | | | | | | | |
| 645 | C | | | | $A1^{+}[P1_{<G>}]$ | | | C.V. / coût |
| | | | | | $A2^{+}[P1_{<G>}]$ | | | C.V. / coût |
| | | | | $P2_{<G>}$ | | | | |
| 646 | M | $P3_{<G>} \ldots$ | | | | | | |
| 647 | C | | | $\ldots P3_{<G>}$ | | | | C.V. / coût |
| 648 | L | | | | | | | |
| 649 | C | | | | $\{-\}P3_{<G>}$ | | | |
| 650 | ? | | | | | | | |
| 651 | M | | | | | | | |
| 652 | L | | | | | $(?)_{<G>}$ | | |
| 653 | M | | | | | | | |
| 654 | L | | | | | $P4_{<archis>}$: | $(A3^{+}[P3])_{<G>}$ | C.V. / Fonctionnel |
| | | | | | | | $\{-\}A3_{<G>}$ | |
| | | | | | | | $A4^{+}P4[=$ | C.V./ Utilisation |
| | | | | | | | $\neg P3]_{<archis>}$ | |

Légende :

C.V. : circulation verticale
$P1_{<G>}$ : diminution des ascenseurs
A1 : baisser les coûts d'aménagement
A2 : baisser les coûts d'entretien
$P2_{<G>}[=precisionP1_{<G>}]$ : suppression ascenseur restaurant
$P3_{<G>}[=precision/designation P1_{<G>}]$
$P3_{<G>}$ : suppression ascenseur près de l'accueil
Pas d'accord avec dimunition de l'ascenseur accueil $P3_{<G>}$
A3 : peut fonctionner avec un seul ascenseur ou monte charge
{−}A3 pas d'accord avec A3 pour les architectes
A4 : accueil privilégié
P4 : ascenseur client et un accueil privilégié (plan archis) = non (¬) P3

La séquence commence avec une sous-séquence de reconstruction « degré zéro » (lecture à haute voix du fax) par C, responsable de ce projet au sein du groupe des architectes, du point de vue de G (diminuer le nombre d'ascenseurs, et en particulier, celui à côté du restaurant), avec les arguments qui y sont associés (baisser les coûts d'aménagement et d'entretien). Cette phase est terminée par la co-construction de M et C (lignes 646, 647) d'une précision de l'ascenseur dont il est question, en désignant sur le plan (l'ascenseur près de l'accueil).

Tout naturellement, une fois que le point de vue de G est mutuellement compris (la participation de M aux dires de C), C émet une évaluation négative sur le point de vue de G (ligne 649) : « c'est là où il y a une confusion ». Les bases de l'interaction argumentative, basculant sur le point de vue des architectes eux-mêmes, sont maintenant mises en place : il y a évaluation négative de la proposition argumentée de l'adversaire absent, suffisamment comprise dans ce contexte (Clark & Schaefer, 1989).

Dans 654 le troisième participant, L, entre en jeu, tout d'abord, en complétant l'énoncé inachevé de C (649. « parce que pour lui [G] …») et en poursuivant son propre énoncé (652. L. « oui (4.0) enfin d'après lui moi [parce que …»). En ce faisant, il achève la reconstruction du point de vue de G, en le mettant en relation avec le point de vue argumenté (« accueil privilégié ») des architectes. Ainsi, la double *problématisation* est complète, « **lui** » et « **nous** » sont mis en relation :

```
654. L. moi lui il pense que:: on peut fonctionner avec un seul euh ascenseur ou
     mont-charge ici (..) euh pour l'instant en fait nous ce qu'on avait prévu c'était
     (qu' t'avais) un accueil ici un petit peu privilégié avec un ascenseur client /
```

En résumé, lors de cette courte séquence, nous avons analysé la circulation des activités cognitives de conception entre les participants, et la collaboration dans ces activités. L'élaboration successive et la mise en relation des points de vue argumentés des participants « absents » et « présents », avec l'expression des évaluations, permettent la mise en place d'une situation d'argumentation délibérative, qui sera à analyser par la suite.

Il est à noter que, si cette présentation est conçue pour montrer la distribution de l'activité au travers les participants, d'autres présentations des mêmes résultats d'analyse seraient utiles, par exemple, sous la forme des colonnes qui montrent comment le raisonnement collectif « circule » à travers les différentes dimensions du raisonnement architectural (« circulation horizontale », « circulation verticale », « fonctionnalité et espace »).

## 5   CONCLUSION

Notre démarche à l'analyse d'une situation de re-conception collective en architecture met l'accent sur la nécessité pour les participants de reconstruire des points de vue des présents et des absents, afin de pouvoir engager une argumentation délibérative. Bien que ces processus puissent paraître bien spécifiques à notre situation d'étude, nous postulons qu'ils seraient mis en œuvre dans toute situation de conception de produits qui exige la prise en compte des exigences d'un commanditaire externe au groupe de concepteurs. Les analyses en cours, partants de la méthode ébauchée ici, et illustrées sur une séquence préliminaire à l'argumentation même, s'attachent à la description fine de l'interaction entre les processus argumentatifs et l'élaboration collective de nouvelles solutions.